\documentclass[conference]{IEEEtran}
\IEEEoverridecommandlockouts
\usepackage{cite}
\usepackage{booktabs}
\usepackage{amsmath,amssymb,amsfonts}
\usepackage{algorithmic}
\usepackage{graphicx}
\usepackage{textcomp}
\usepackage{xcolor}
\usepackage{adjustbox}
\usepackage{tikz}
\usetikzlibrary{positioning}

\def\BibTeX{{\rm B\kern-.05em{\sc i\kern-.025em b}\kern-.08em
    T\kern-.1667em\lower.7ex\hbox{E}\kern-.125emX}}
\begin{document}

\title{Speech Quality Factors for Traditional and Neural-Based Low Bit Rate Vocoders
}

 \author{\IEEEauthorblockN{Wissam A. Jassim}
 \IEEEauthorblockA{\textit{School of Computer Science} \\
 \textit{University College Dublin}\\
 Dublin, Ireland \\
 wissam.a.jassim@gmail.com}
 \and
 \IEEEauthorblockN{Jan Skoglund}
 \IEEEauthorblockA{\textit{Chrome Media} \\
 \textit{Google}\\
 San Francisco, CA, USA\\
 jks@google.com}
 \and
 \IEEEauthorblockN{Michael Chinen}
 \IEEEauthorblockA{\textit{Chrome Media } \\
 \textit{Google}\\
 San Francisco, CA, USA \\
mchinen@google.com}
 \and
 \IEEEauthorblockN{Andrew Hines}
 \IEEEauthorblockA{\textit{School of Computer Science} \\
 \textit{University College Dublin}\\
 Dublin, Ireland \\
 andrew.hines@ucd.ie}

}

\IEEEpubid{\makebox[\columnwidth]{978-1-7281-5965-2/20/\$31.00 
\copyright 2020 IEEE \hfill} 
\hspace{\columnsep}\makebox[\columnwidth]{ }}
\maketitle

\begin{abstract}
This study compares the performances of different algorithms for coding speech at low bit rates. In addition to widely deployed traditional vocoders, a selection of recently developed generative-model-based coders at different bit rates are contrasted. Performance analysis of the coded speech is evaluated for different quality aspects: accuracy of pitch periods estimation, the word error rates for automatic speech recognition, and the influence of speaker gender and coding delays. A number of performance metrics of speech samples taken from a publicly available database were compared with subjective scores. Results from subjective quality assessment do not correlate well with existing full reference speech quality metrics. The results provide valuable insights into aspects of the speech signal that will be used to develop a novel metric to accurately predict speech quality from generative-model-based coders. 
\end{abstract}



\section{Introduction}
A vocoder system refers to the process that analyzes and synthesizes human speech for a wide range of applications such as speech compression, voice transformation, and multiplexing. It has been studied for decades and many well-developed algorithms have been introduced \cite{1162554}\cite{doi:10.1121/1.1912679}. These algorithms are designed to compress speech signals with low bit rates and high speech quality. The performance of any vocoder system is determined by other desirable properties such as robustness to different speakers/languages, robustness to channel errors, computational complexity and memory aspects, and minimizing the coding delay \cite{10.5555/546671}. 

Recently, several deep neural network (DNN) speech synthesis algorithms that contain a generative model have been proposed for text-to-speech synthesis and low bit-rate coding (compression)  \cite{8461368}\cite{DBLP:journals/corr/ArikDGMPPRZ17}\cite{Kleijn2017WavenetBL}\cite{prenger2019waveglow}. WaveNet \cite{45774}, a deep neural network for generating raw audio waveforms, was the first method of these algorithms. It was originally designed for text-to-speech synthesis and has since been used for parametric coding~\cite{Kleijn2017WavenetBL}, in which the speech signal is represented as a sequence of parameters extracted at the encoder. 
To improve memory efficiency, SampleRNN~\cite{Mehri2016SampleRNNAU} uses a sparse recurrent neural network (RNN) instead of the convolutional networks. It has been shown to be useful for speech coding \cite{8682435} as it is able to overcome the problem of
modeling extremely high-resolution temporal data. WaveRNN \cite{DBLP:journals/corr/abs-1802-08435} also focused on finding more efficient models in order to reduce the complexity of speech synthesis compared to WaveNet. Furthermore, the LPCNet model \cite{8682804} has shown be able to provide lower complexity and real-time operation by combining linear prediction with WaveRNN. It achieved significantly higher quality than WaveRNN for the same network size. All of the models generate samples by sampling from a distribution, and most are autoregressive.  This means the number of waveforms that are likely for a given sequence of conditioning input increases with the model's uncertainty. The typical errors generated by model failures are incorrect mappings on the speech manifold, such as phoneme mismatches or slurred speech, as opposed to non-speech artifacts introduced by traditional coders.

The neural-based generative networks have recently been adopted to improve the performance of other types of coders (waveform and parametric coders). For example, a backward-compatible way of improving the quality of the low bit rate Opus coder \cite{rfc6716} by re-synthesizing speech from the decoded parameters using two different neural generative models, WaveNet (high complexity, and high-latency architecture) and LPCNet (low-complexity,
low-latency RNN-based generative model) was proposed in \cite{Skoglund2019ImprovingOL}. The two systems were used to extract conditioning features from the Opus bit stream coded at 6 kb/s. These conditioning features represent the spectral shape and the pitch of the signal for performing signal decoding and reconstruction. According to the subjective tests, synthesized speech using LPCNet outperformed the output of the standard Opus decoder for the same 6 kb/s bit stream.      

The speech quality assessments summarised in Section~\ref{sec:datasets} rated the neural-based coders highly. However, in Section~\ref{sec:pesqtest} we observe that traditional full reference objective speech quality metrics such as PESQ (Perceptual Evaluation of Speech Quality) measure~\cite{pesq_itu} fail to provide accurate quality scores for the speech signals processed by this kind of speech coders. Therefore, a new quality measure is needed that is robust enough to work with different types of coders. As a first step, making meaningful comparisons between the performance of the most recent neural-based coders is necessary to determine the related factors that need to be taken into account when designing a new quality measures. 

\begin{table*}[t]
  \centering
  \vspace{-4mm}
  \caption{Coders adopted in this study.}
    \begin{adjustbox}{width=\textwidth}
    \begin{tabular}{lllllll}
    Abbreviation & Coder Name & Bit rate & Type & Database & Notes & Reference \\
    \midrule
    \midrule
    LPCNetUnquant & Unquantized LPCNet & ---  & Generative-model-based coder & Set 1 \cite{Valin2019} & LPCNet operating on Opus unquantized features & \cite{Valin2019} \\
    \midrule
    Opus9.0 & Opus & 9~kb/s & Hybrid waveform-maching coder & Set 1 \cite{Valin2019}, Set 2 \cite{Skoglund2019ImprovingOL} & Wideband vocoder (SILK mode)  & \cite{rfc6716} \cite{Skoglund2019ImprovingOL} \\
    \midrule
    WaveNet6.0 & WaveNet & 6~kb/s & Generative-model-based coder & Set 2 \cite{Skoglund2019ImprovingOL} & WaveNet operating on Opus quantized features & \cite{Skoglund2019ImprovingOL} \\
    \midrule
    LPCNet1.6 & LPCNet & 1.6~kb/s & Generative-model-based coder & Set 1 \cite{Valin2019} & WaveRNN + linear prediction & \cite{8682804}\cite{Valin2019} \\
    \midrule
    LPCNet6.0 & LPCNet & 6~kb/s & Generative-model-based coder & Set 2 \cite{Skoglund2019ImprovingOL} & LPCNet operating on Opus quantized features & \cite{Skoglund2019ImprovingOL} \\
    \midrule
    MELP2.4 & MELP & 2.4~kb/s & Source-filter coder & Set 1 \cite{Valin2019} & Narrowband vocoder & \cite{540325} \\
    \midrule
    Opus6.0 & Opus & 6~kb/s & Hybrid waveform-matching coder & Set 2 \cite{Skoglund2019ImprovingOL} & Narrowband vocoder (SILK mode)  & \cite{rfc6716} \cite{Skoglund2019ImprovingOL} \\
    \midrule
    Speex4.0 & Speex & 4~kb/s & Hybrid waveform coder? & Set 1 \cite{Valin2019}, Set 2 \cite{Skoglund2019ImprovingOL} & Wideband vocoder (wideband quality 0) & \cite{speex2007} \\
    \bottomrule
    \end{tabular}%
    \end{adjustbox}
  \label{tab_1}%
    \vspace{-4mm}
\end{table*}%

Consequently, in Section~\ref{sec:methods} of this study, the quality of speech coders is evaluated in terms of multi-dimensional aspects. The coded speech signals obtained from most recent coders are evaluated in terms of the following quality aspects: subjective listening scores with a MUSHRA-inspired methodology~\cite{2003RECOMMENDATIONIB}, accuracy of pitch period estimation, word error rate for automatic speech recognition, robustness to speaker gender, and robustness to coding delays.          

\section{Speech Quality Datasets for Coded Speech}
\label{sec:datasets}

In this study, we employed the same speech materials with their subjective listening scores that were used to test the quality of the LPCNet coder at 1.6~kb/s in \cite{Valin2019} and the modified Opus coder at 6~kb/s in \cite{Skoglund2019ImprovingOL}. As stated in \cite{Valin2019}, 16 samples from 3 male and 3 female speakers (Set 1) were used in the testing stage. On the other hand, the speech database employed to test the quality of the modified Opus coder in  \cite{Skoglund2019ImprovingOL} consists of 16 samples from 3 male and 3 female speakers (Set 2). For the two studies, all samples are part of the NTT Multi-Lingual Speech Database for Telephonometry database \cite{NTT_database}. In addition, all samples from the selected speakers for the test were excluded from the training set. The subjective listening tests with a MUSHRA-inspired methodology \cite{2003RECOMMENDATIONIB} were conducted on 100 listeners for each test.

\begin{figure}[ht!]
\centerline{\includegraphics[width=.7\linewidth]{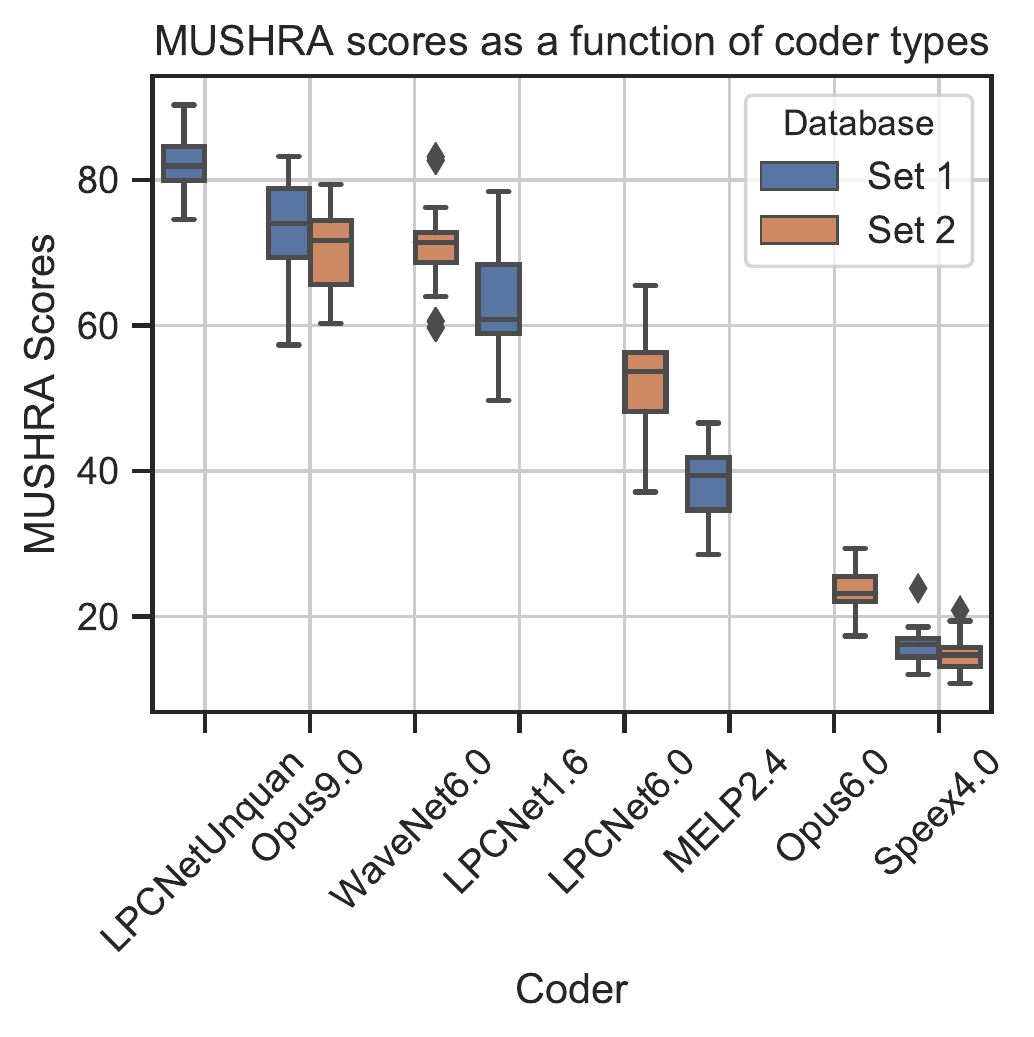}}
\vspace{-5mm}
\caption{Subjective listening results of different types of coders taken from the LPCNet \cite{Valin2019} and OpusNet \cite{Skoglund2019ImprovingOL} studies. Note that the 9 kb/s Opus and 4 kb/s Speex coders were used in both studies. }
\label{fig1}
\vspace{-3mm}
\end{figure}

Table~\ref{tab_1} lists the speech coders that were employed for performance analysis in this study. The coders are sorted in descending order according to their MUSHRA scores. Fig.~\ref{fig1} shows these scores as a function of coder type. It can be seen that the best results were obtained by the LPCNet coder operating on unquantized features (LPCNetUnquan). The 9~kb/s Opus (Opus9.0) and 6~kb/s WaveNet (WaveNet6.0) coders achieved comparable results. The RNN-based coder at 1.6~kb/s (LPCNet1.6) outperformed the same coder operating on Opus quantized features at 6~kb/s (LPCNet6.0). Both WaveNet6.0 and LPCNet6.0 at provided higher quality results than that of the standard Opus decoder for the same 6~kb/s bit stream. The quality of MELP coder (source-filter coder) at 2.4~kb/s obtained better results than that of the Speex coder at 4~kb/s.   

\begin{figure}[ht!]
\centerline{\includegraphics[width=.69\linewidth]{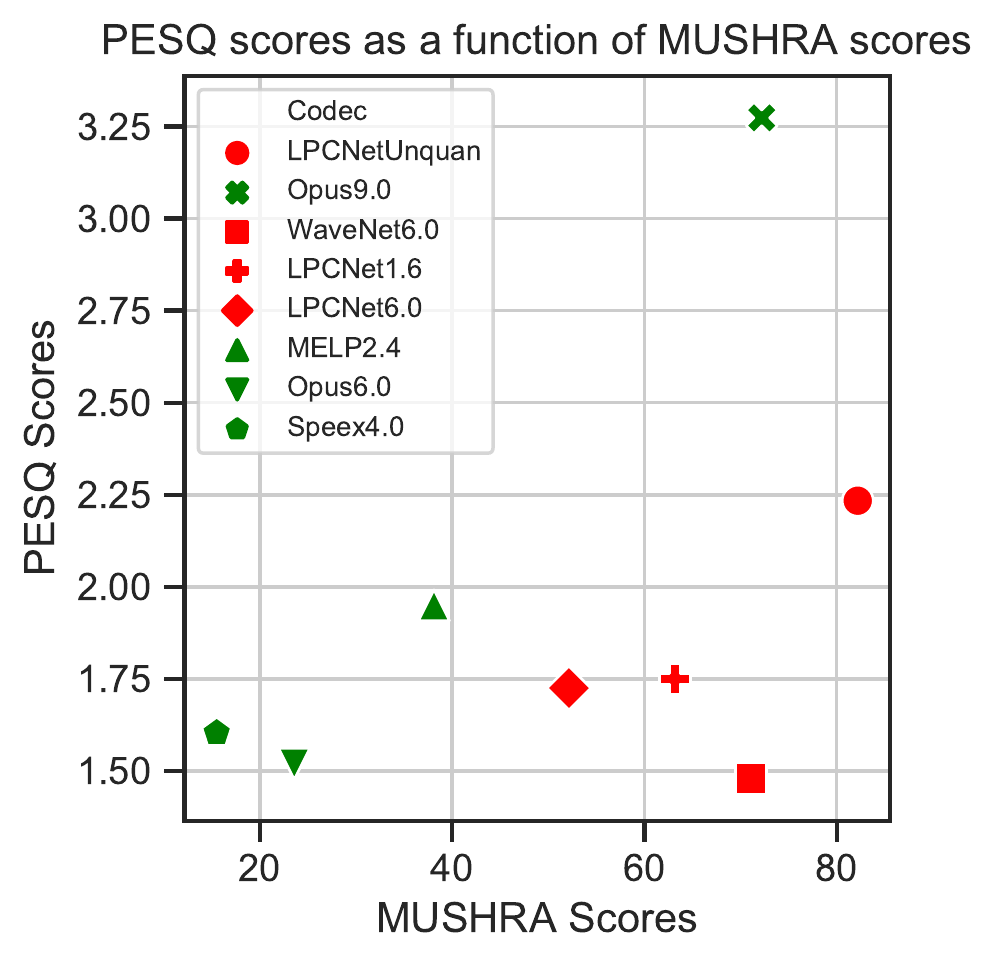}}
\vspace{-4mm}
\caption{Objective PESQ measure against subjective MUSHRA scores for different coders. Results were averaged by coder type and compared to the corresponding average MUSHRA scores. }
\label{fig2}
\vspace{-4mm}
\end{figure}

\begin{figure*}[t!h]
\centerline{\includegraphics[width=.7\linewidth]{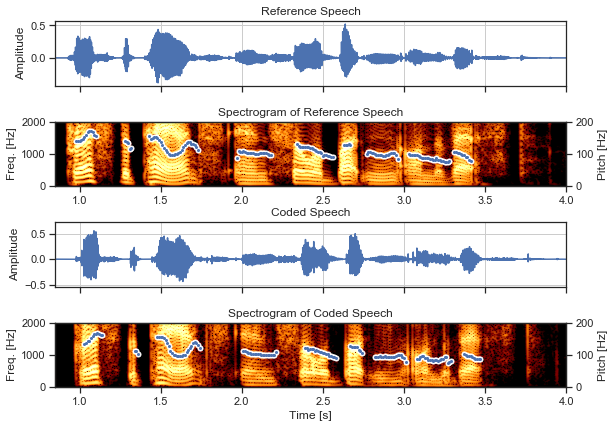}}
\vspace{-5mm}
\caption{Example of pitch (fundamental frequency) plots for a reference speech sample with its corresponding synthesised sample. In this example the speech signal was processed by the LPCNet6.0 coder. The spectrogram frequencies are shown up to 2 kHz only.}
\label{fig_pitch}
\vspace{-4mm}
\end{figure*}


\section{Objective Speech Quality Evaluation}
\label{sec:pesqtest}
In order to evaluate the quality of the speech signals coded by different coders, PESQ \cite{pesq_itu} scores were computed for the reference speech samples and their corresponding coded (synthesized) samples. Results were averaged by coder type and compared to the average MUSHRA scores per coder type. Fig.~\ref{fig2} shows the PESQ scores as a function of MUSHRA scores. In the plot, the relationship points are highlighted in two different colors. The points highlighted in green represent the results for the parametric coders (Opus9.0, MELP2.4, Opus6.0 and Speex4.0) whereas the points highlighted in red are the results for the generative-model-based coders (LPCNetUnquan, WaveNet6.0, LPCNet1.6 and LPCNet6.0). It can be seen that the PESQ scores of the parametric coders obtained good correlation with the subjective listening results. On the other hand, the generative-model-based coders provided poor correlation in terms of PESQ against MUSHRA scores relationship. This suggests that the traditional speech quality metrics may not be suitable for evaluating the performance of generative-model-based coders. Standard objective speech quality measures such as PESQ are unable to reflect the distortions in the speech signals of these coders. Similar behaviour was observed by the authors with other full reference quality metrics (e.g. POLQA~\cite{polqa3} and ViSQOL~\cite{hines2015visqol}) motivating this study and more comprehensive experiments to design a new quality measure for these kind of coders. 

\section{Comparative Analysis of Coder Quality}
\label{sec:methods}
It is important to investigate how speech coders, especially the generative-model-based coders, behave in terms of other quality aspects as illustrated in the following subsections. Having an idea about the effect of these aspects on coded speech, it is useful to improve the performance of the current coders and to develop more robust coders in the future.   

\subsection{Accuracy of pitch estimation}
As an example of how the pitch (fundamental frequency) looks like, Fig.~\ref{fig_pitch} shows the plots of the pitch sequence as a function of time for a reference speech sample taken from the NTT database with its corresponding speech obtained from a generative-model-based coder. The pitch frequencies were computed using Praat software \cite{praat2001}. The spectrogram representation of the two signals are also shown in Fig.~\ref{fig_pitch}.

\begin{figure}[t!]
\centerline{\includegraphics[width=.69\linewidth]{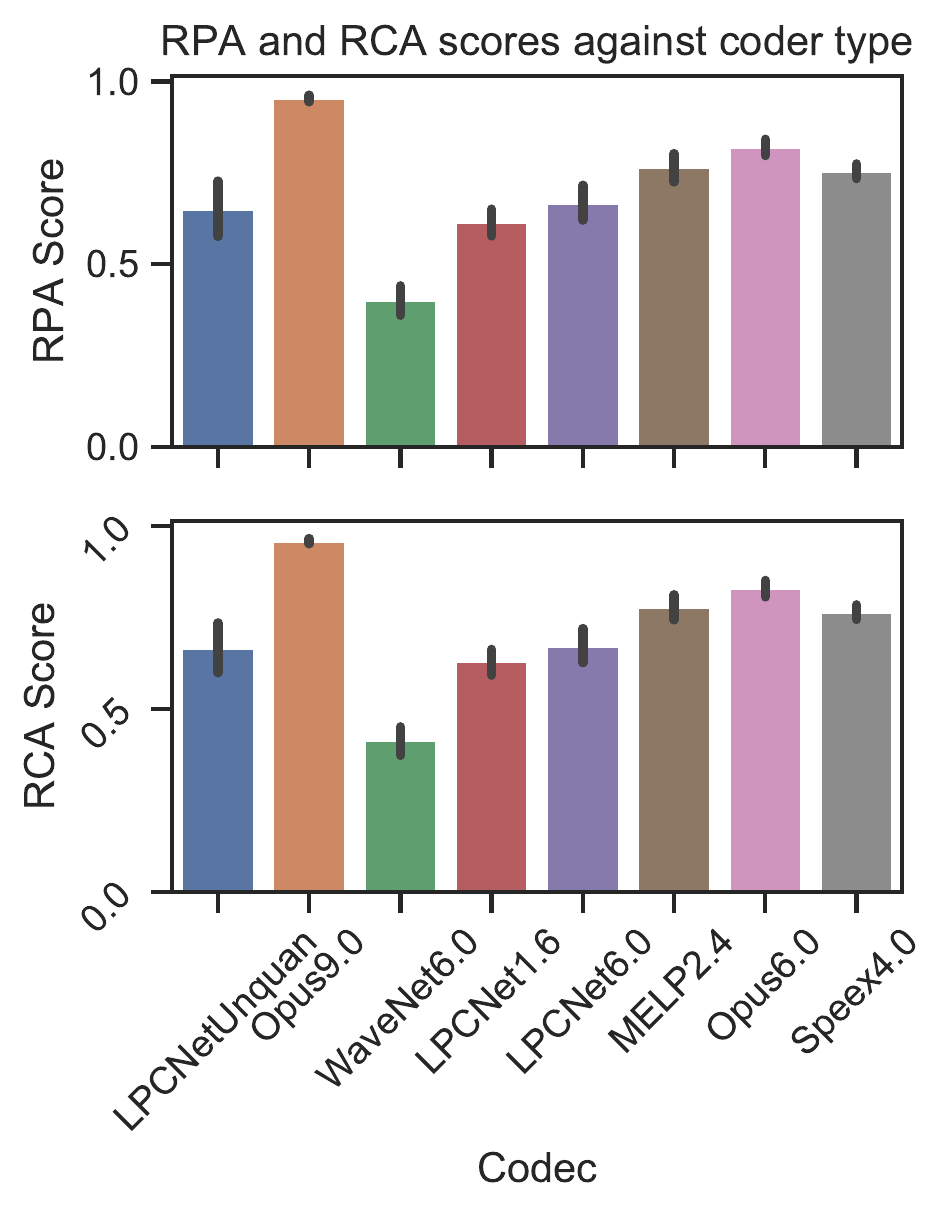}}
\vspace{-4mm}
\caption{Pitch accuracy scores as a function of coder types.}
\vspace{-4mm}
\label{fig_rpa_rca}
\end{figure}

\begin{figure}[t!]
\centerline{\includegraphics[width=.69\linewidth]{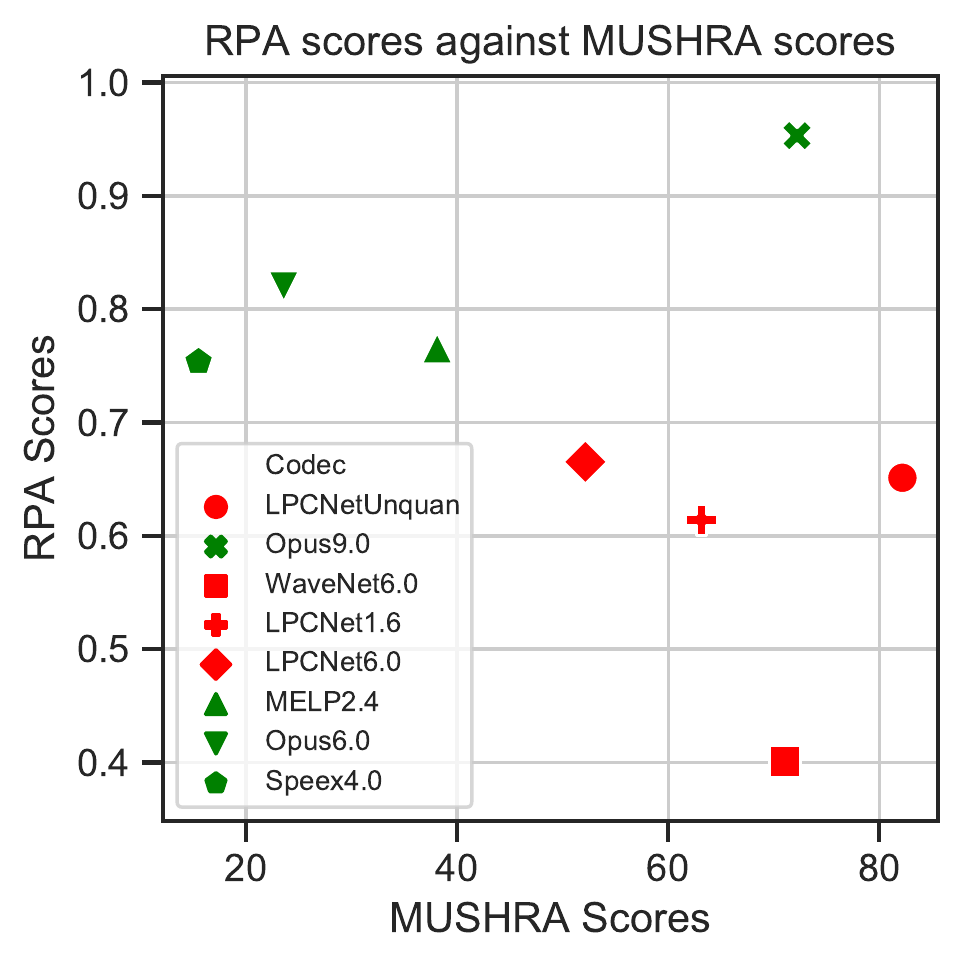}}
\vspace{-4mm}
\caption{Pitch accuracy scores against subjective MUSHRA scores for different coders. Results were averaged by coder type and compared to the corresponding average MUSHRA scores. }
\label{fig_rpa_mushra}
\vspace{-4mm}
\end{figure}

The $\text{mir}\_\text{eval}$ \cite{RaffelMHSNLE14} python package was used to compute pitch accuracy given two pitch sequences. The results are compared in terms of two measures \cite{4156215}~\cite{460}: raw pitch accuracy (RPA) and raw chroma accuracy (RCA) with a threshold of 50 cents. Fig.~\ref{fig_rpa_rca} shows the RPA and RCA results against coder type. It can be seen that the best results were obtained by the Opus coder at 9 kbps. The other parametric coders (MELP2.4, Opus6.0 and Speex4.0) also outperformed by the generative-model-based coders. Among these coders, the LPCNetUnquan and LPCNet6.0 achieved comparable accuracies in terms of pitch estimation, whereas the WaveNet6.0 obtained the lowest RPA and RCA scores.
Furthermore, the RPA results averaged by coder were plotted as a function of the average MUSHRA scores as shown in Fig.~\ref{fig_rpa_mushra}. Again, the neural-based coders obtained poor correlation compared to that of the other coders. 

\subsection{Word error rates for automatic speech recognition}
\begin{figure}[t!]
\centerline{\includegraphics[width=.69\linewidth]{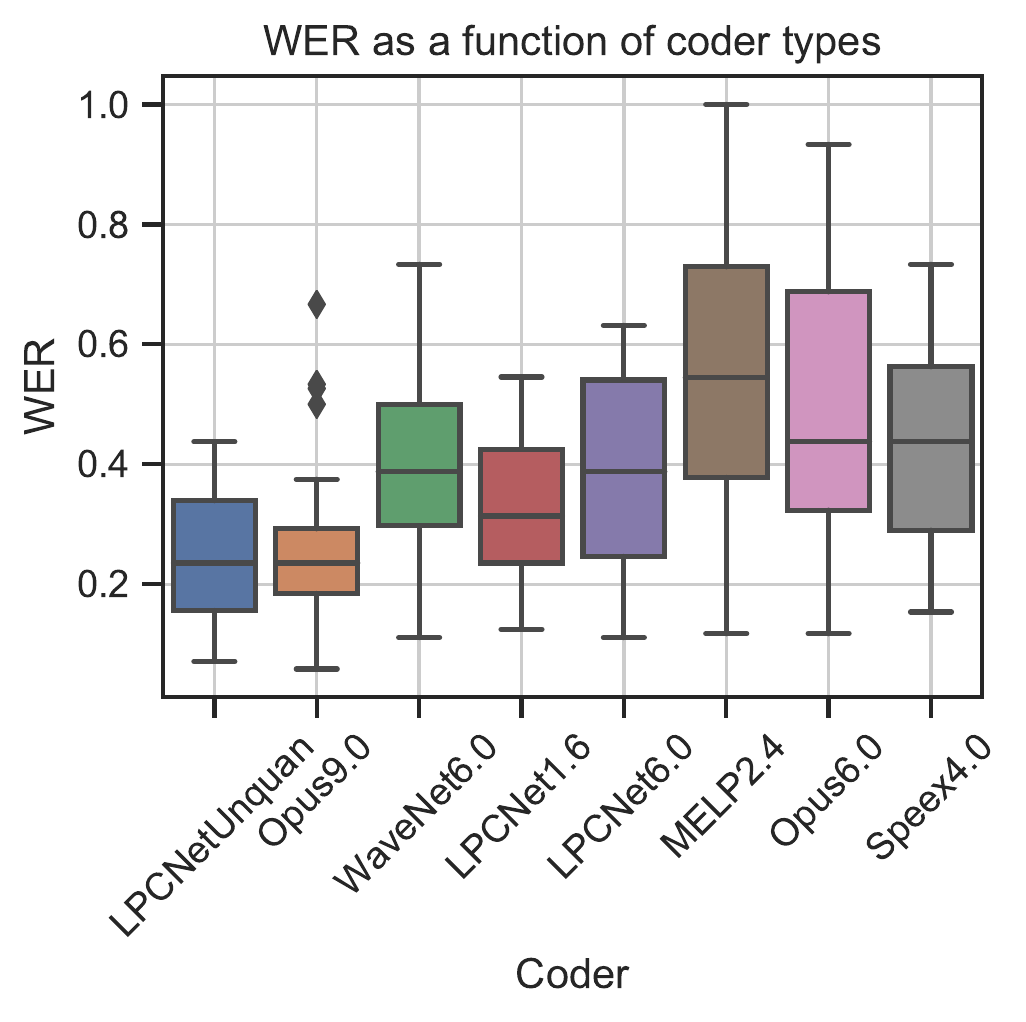}}
\vspace{-4mm}
\caption{Results of WER as a distance measure between the ground-truth transcript and the hypothesis transcript.}
\label{fig_wer}
\vspace{-4mm}
\end{figure}

\begin{figure}[ht!]
\centerline{\includegraphics[width=.69\linewidth]{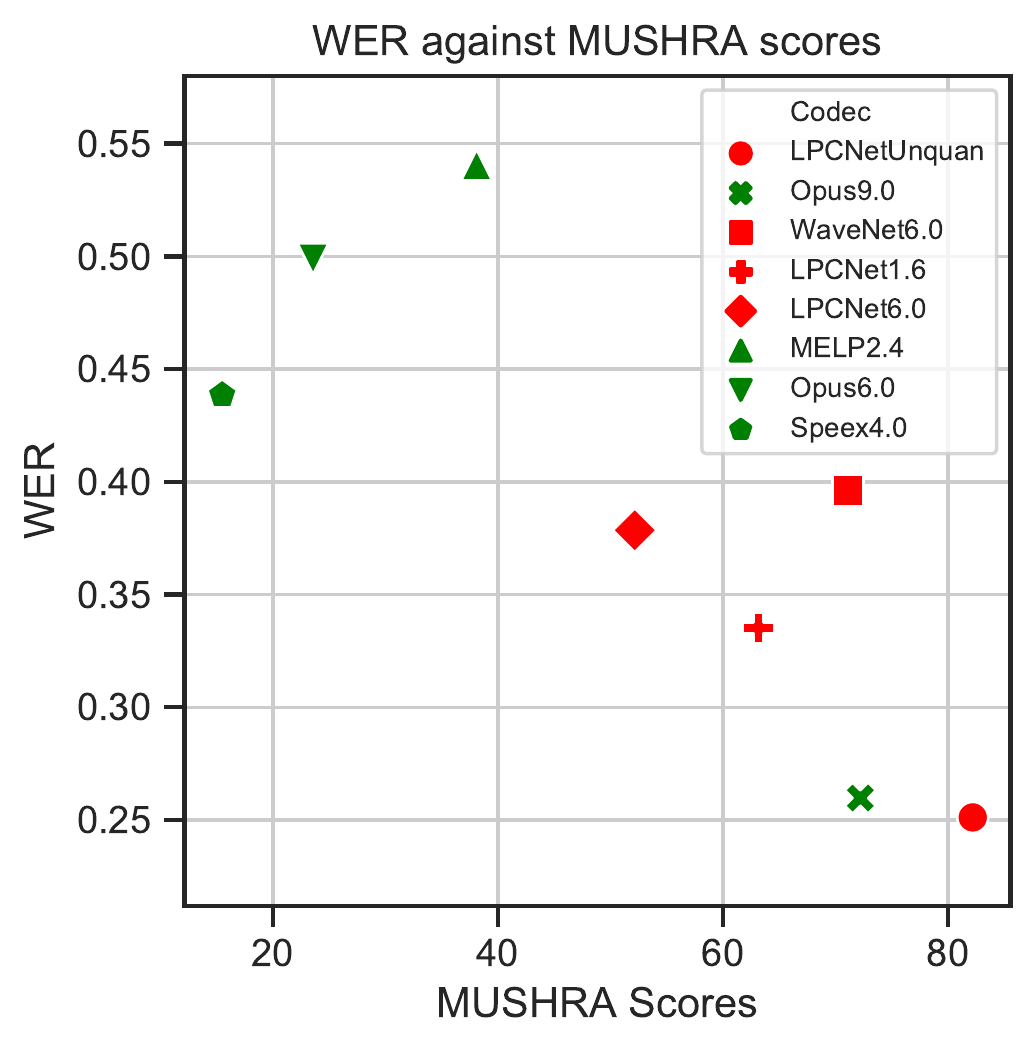}}
\caption{WER against subjective MUSHRA scores for different coders. Results were averaged by coder type and compared to the corresponding average MUSHRA scores.}
\label{fig_wer_mushra}
\vspace{-4mm}
\end{figure}

In this analysis, the speech signal obtained from each speech coder was fed as an input to the speech recognition library~\cite{zhang_2017} with the Google speech-to-text API \cite{speechapi}. The word error rate (WER) between the ground-truth transcript and the hypothesis transcript of the API was then computed. Fig.~\ref{fig_wer} shows the plot of WER results for all coders. As expected, the Opus9.0 and LPCNetUnquan coders achieved the best results. The neural-based coders at low bit rates (WaveNet6.0, LPCNet1.6 and LPCNet6.0) outperformed the traditional parametric coders (MELP2.4, Opus6.0 and Speex4.0). The MELP2.4 coder obtained the worst error rates whereas the Opus6.0 and Speex4.0 coders achieved comparable results. The WER versus MUSHRA scores relationship is shown in Fig.~\ref{fig_wer_mushra}. The overall trend is an inverse correlation of WER to quality with the parametric coders (except Opus9.0) displaying higher WERs than the generative-model-based coders. 

The pitch tracking results highlighted that while there are differences in pitch tracking observed from the neural-based coders, these do not influence perceived quality and a new objective quality metric needs to take this into account. From the WER results, a similar trend is observed for intelligibility. Fig.~\ref{fig_wer_mushra} highlights that there is a strong inverse correlation between WER and the MUSHRA scores, i.e. WER is a better predictor of quality than the objective quality metric results seen in Fig.~\ref{fig2}.  This points to the potential to use utomatic speech recognition (ASR) as part of an objective quality metric solution although as the ASR model is language dependant, any such solution would be less flexible than the current objective speech quality models that are not language specific in design.

\subsection{Robustness to speaker's gender}
In this analysis, we investigate the influence of speaker’s gender on the quality of coded speech. Before describing the outcomes of this experiment, it is necessary to mention that each speech sample in the employed database contains two utterances, and each utterance was spoken by a different speaker. This approach generated speech samples with four categories of speaker's gender: Male-Male, Male-Female, Female-Male and Female-Female. The Male-Male category means that the two utterances in the sample were spoken by two male speakers, whereas female-female category contains speech samples that their utterances were spoken by two female speakers, and so on for other categories.

\begin{figure*}[t!]
\centerline{\includegraphics[width=.7\linewidth]{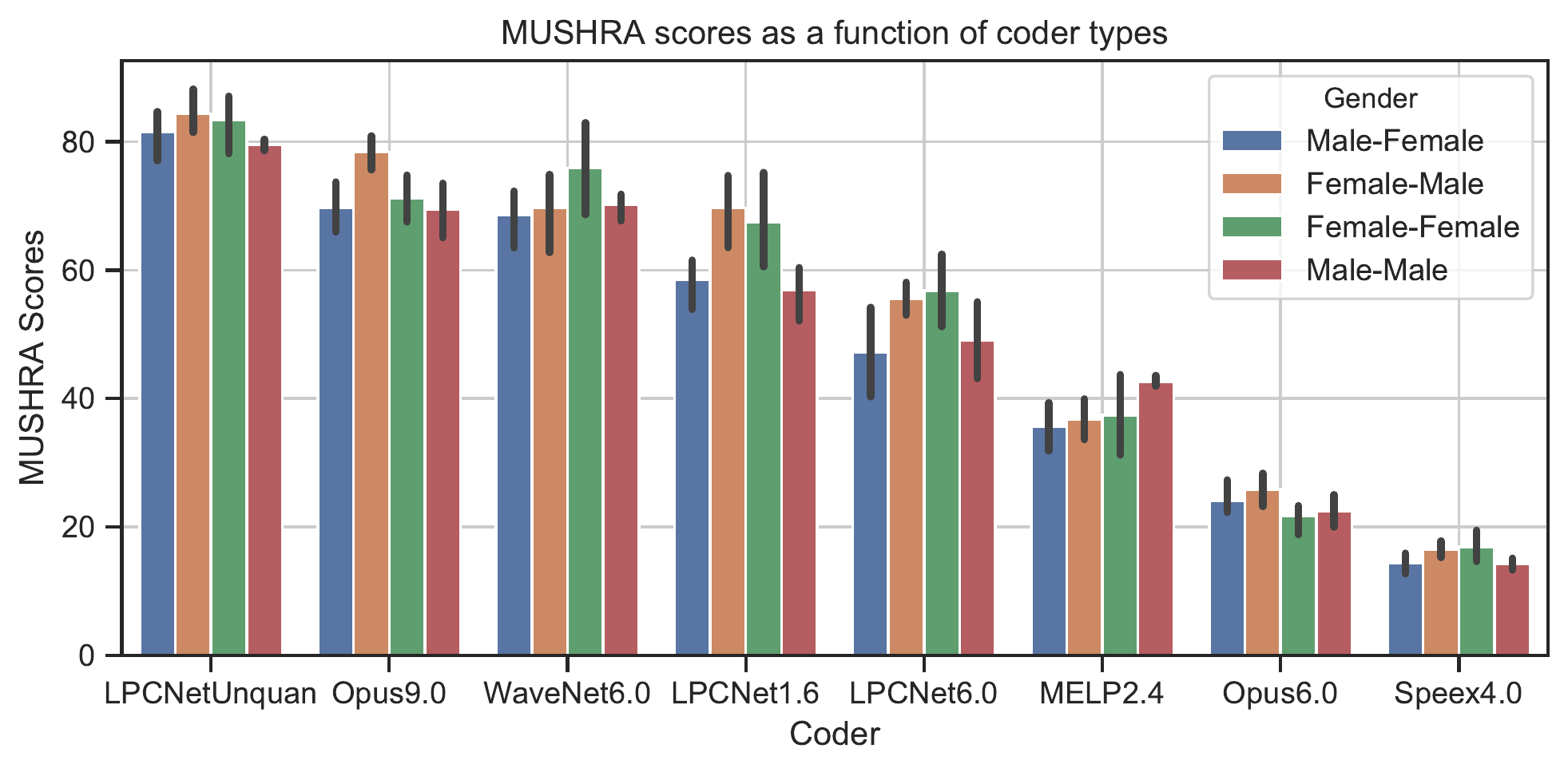}}
\vspace{-2.5mm}
\caption{Subjective MUSHRA scores for different coders and speaker's gender category}
\label{gender_effect}
\vspace{-4mm}
\end{figure*}

The subjective listening results were reported as a function of speaker's gender for each coder, and the results are shown in Fig.~\ref{gender_effect}. The speech samples with Female-Male category achieved highest subjective scores for Opus coders. For the generative-model-based coders, the highest subjective scores were obtained when the first utterance of the speech sample is spoken by a female speaker, i.e. the speech samples with Female-Male and Female-Female categories. This behaviour is consistent with that of the Speex4.0 coder. The Male-Male speech samples achieved the best results for the MELP2.4 coder. It can be seen that the speaker's gender has an impact on the overall subjective listening results as the Mushra scores seemed to be affected by speaker's gender. This factor needs to be further investigated when designing new speech coders especially for the neural-based models.

\begin{figure}[t!]
\centerline{\includegraphics[width=.6\linewidth]{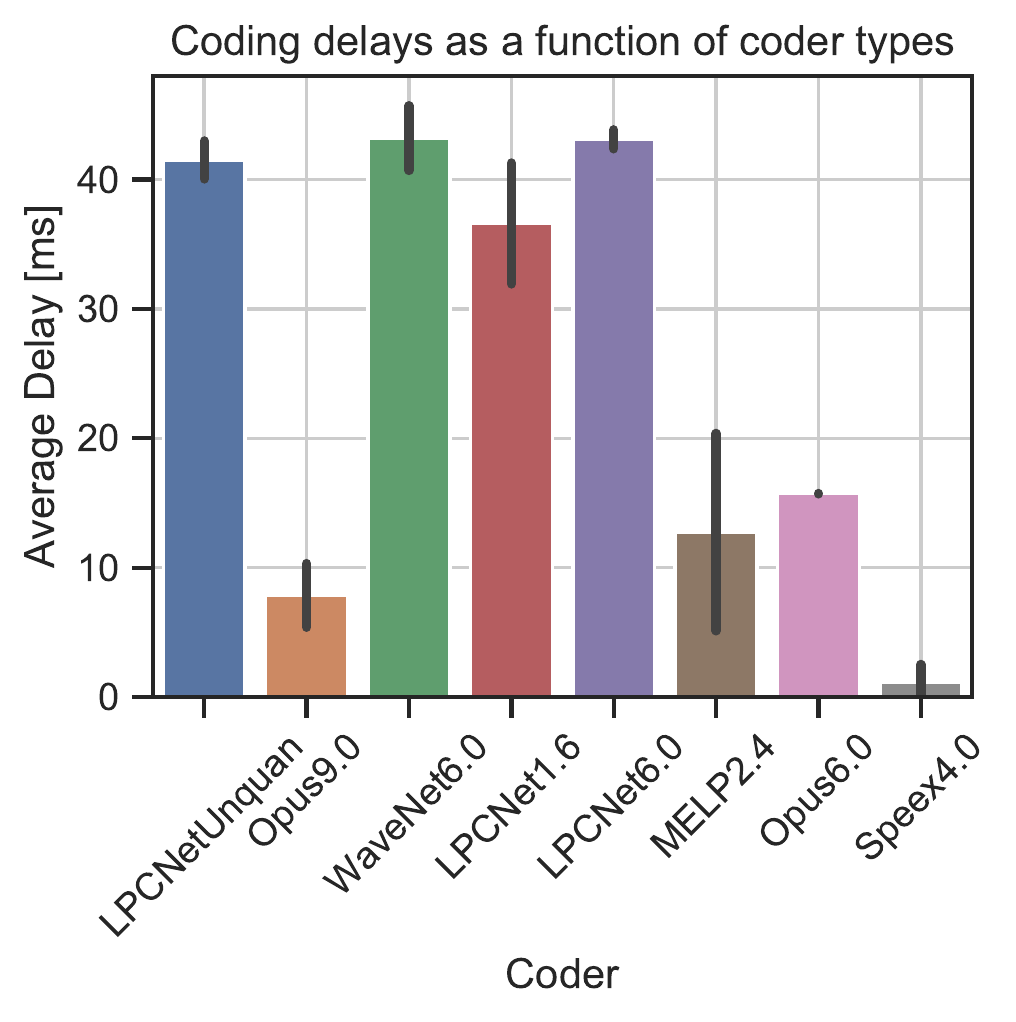}}
\vspace{-4mm}
\caption{Results of average coding delays between reference and coded speech samples.}
\label{avg_delays}
\vspace{-4mm}
\end{figure}

\subsection{Robustness to coding delays}
\begin{figure}[ht!]
\centerline{\includegraphics[width=.7\linewidth]{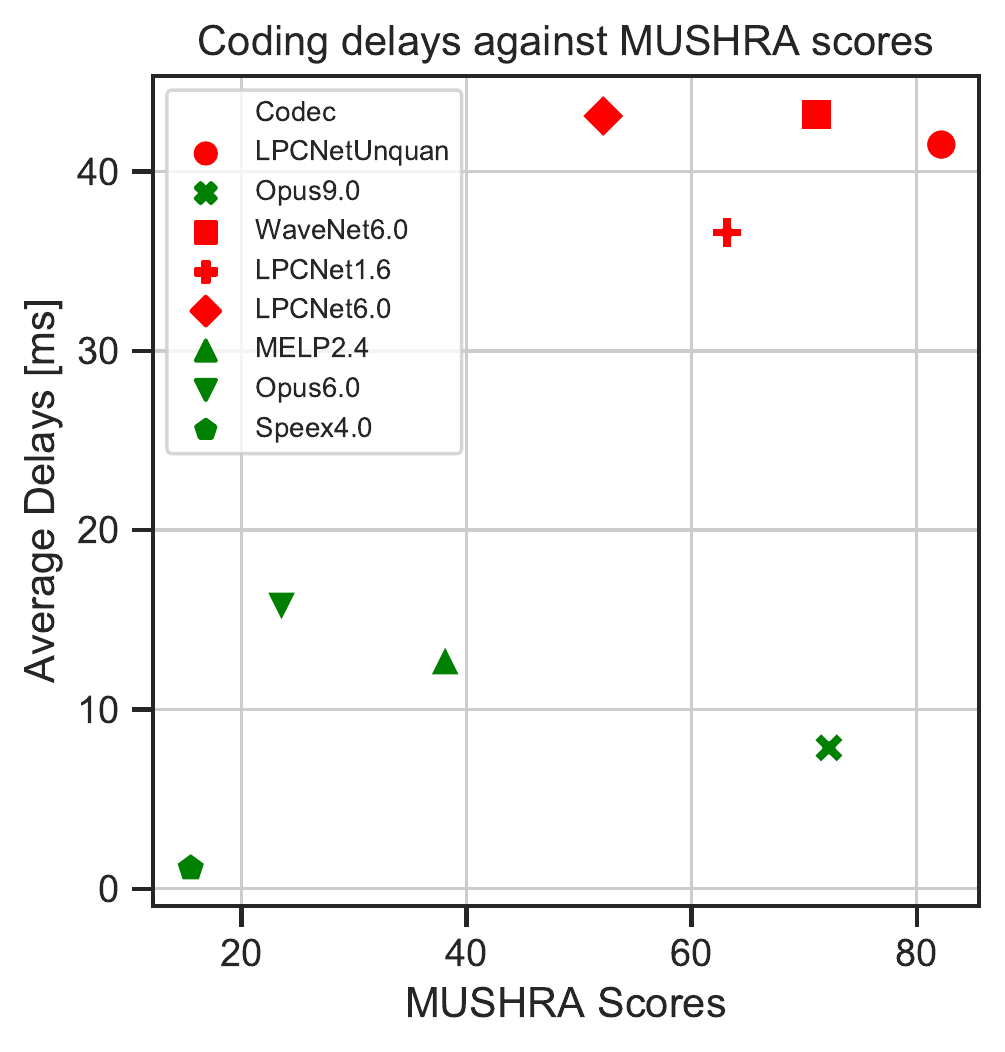}}
\caption{Average coding delays against subjective MUSHRA scores for different coders. Results were averaged by coder type and compared to the corresponding average MUSHRA scores.}
\label{avg_delays_mushra}
\vspace{-4mm}
\end{figure}

The effect of coding delay on the quality of different speech coders was also tested in this study. The cross-correlation was computed between the reference and coded speech signals, and the index of the maximum values (argmax) was taken as time leads or lags. Fig.~\ref{avg_delays} shows the average coding delays as a function of coder type. As expected, the generative-model-based coders incur longer delays in time compared to that of the other coders. For delays plotted against the subjective scores as shown in Fig.~\ref{avg_delays_mushra} there is some correlation apparent (except Speex4.0). Identifying delay characteristics from the coded speech could be useful for classifying coder behaviour for data driven objective quality models.

\begin{figure*}[ht!]
\centerline{\includegraphics[width=0.9\linewidth]{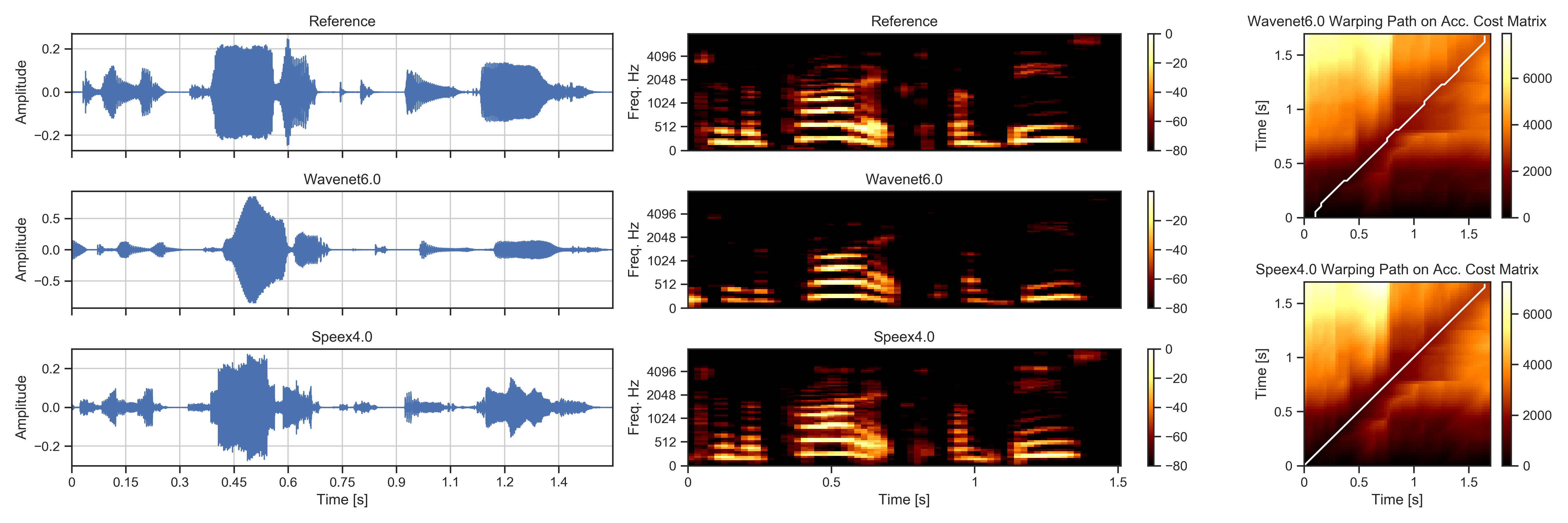}}
\vspace{-5mm}
\caption{Example of 1.5 s of Reference, WaveNet6.0 and Speex4.0 coded speech. The waveforms, Mel spectrogram in dB, and DTW cost matrix plots are shown to highlight the micro-alignment changes in neural-based vocoded speech.}
\label{microdelays}
\vspace{-4mm}
\end{figure*}

Another experiment was run to show the micro-alignment time difference between reference speech and its corresponding coded speech. Fig.~\ref{microdelays} shows waveform of a reference speech sample taken form the same database. The same sample was processed by two coders: the WaveNet6.0 and Speex4.0. A Mel spectrogram with 64 band to 8 kHz was then computed for each signal. The 2 dimensional plot of each Mel spectrogram is shown in the same figure. The dynamic time wrapping (DTW) algorithm was performed on the two Mel spectrograms to measure similarity (in terms of temporal information) between the reference and coded speech waveforms~\cite{muller2015fundamentals}. The optimal path that has the minimal cost (summation of absolute differences) was then tracked and plotted. It can be seen that, although the Speex coder obtained poor subjective scores (see Fig.~\ref{fig1}), it provided very low micro delays as its optimal path is a straight diagonal. On the other hand, the micro-alignment time shifts are very apparent for the WavNet6.0 coder.

\section{Conclusion}
In this study, different types of speech coders were compared in terms of several objective measures. Experiments were conducted using speech samples taken from the NTT database with its subjective MUSHRA scores. The existing full reference speech quality metrics failed to provide scores consistent with subjective quality assessment. Taking into account the influence of the quality aspects considered in this study is necessary to design a new quality metric that is robust enough to work with different coders at different bit rates. One possible approach is to combine results from the objective measures employed in this study based on multi modelling algorithm to design the required speech quality metric.

\section*{Acknowledgments}
This publication has emanated from research supported in part by the Google Chrome University Program and research grants from Science Foundation Ireland (SFI) co-funded under the European Regional Development Fund under Grant Number 13/RC/2289\_P2 and 13/RC/2077.


\vspace{1mm}

\bibliographystyle{IEEEtran}
\bibliography{main}

\end{document}